\documentclass[pdflatex, sn-mathphys]{sn-jnl}

\usepackage[utf8]{inputenc}
\usepackage[T1]{fontenc}

\usepackage{graphicx}% Include figure files
\usepackage{dcolumn}% Align table columns on decimal point
\usepackage{bm}% bold math
\usepackage{siunitx}
\usepackage[nice]{nicefrac}
\usepackage{xcolor}
\usepackage{hyperref}
\usepackage{upgreek}
\usepackage{nameref}

\hypersetup{
    colorlinks,
    linkcolor={black},
    citecolor={blue!50!black},
    urlcolor={blue!80!black}
}

\begin{document}

\title{Direct observation of propagating spin waves in a spin-Hall nano-oscillator}

\author*[1,2]{\fnm{Victor H.} \sur{Gonz\'alez}}\email{vhg23@cam.ac.uk}
\equalcont{These authors contributed equally to this work.} 
\author[3]{\fnm{Frank} \sur{Schulz}}
\equalcont{These authors contributed equally to this work.} 
\author[2,4]{\fnm{Nilamani} \sur{Behera}}
\equalcont{These authors contributed equally to this work.} 
\author[2]{\fnm{Martina} \sur{Ahlberg}}
\author[2,5,6]{\fnm{Akash} \sur{Kumar}}
\author[2]{\fnm{Andreas} \sur{Frisk}}
\author[7]{\fnm{Felix} \sur{Gro{\ss} }}
\author[3]{\fnm{Sven Erik} \sur{Ilse}}
\author[8]{\fnm{Steffen} \sur{Wittrock}}
\author[8]{\fnm{Markus} \sur{Weigand}}
\author[8]{\fnm{Gisela} \sur{Schütz}}
\author*[2,5,6]{\fnm{Johan} \sur{\AA kerman}}\email{johan.akerman@physics.gu.se}
\author*[8]{\fnm{Sebastian} \sur{Wintz}}\email{sebastian.wintz@helmholtz-berlin.de}

\affil[1]{\orgdiv{Department of Applied Mathematics and Theoretical Physics}, \orgname{University of Cambridge}, \postcode{CB3 0EH}, \city{Cambridge}, \country{United Kingdom}}

\affil[2]{\orgdiv{Department of Physics}, \orgname{University of Gothenburg}, \postcode{412 96}, \city{Gothenburg}, \country{Sweden}}

\affil[3]{\orgname{Max-Planck-Institut für Festkörperforschung}, \postcode{705 69}, \city{Stuttgart}, \country{Germany}}

\affil[4]{\orgdiv{Department of Physics, School of Basic Sciences}, \orgname{Indian Institute of Technology Bhubaneswar}, \postcode{752050}, \city{Odisha}, \country{India}}

\affil[5]{\orgdiv{Research Institute of Electrical Communication}, \orgname{Tohoku University}, \postcode{980-8577}, \city{Sendai}, \country{Japan}}

\affil[6]{\orgdiv{Center for Science and Innovation in Spintronics}, \orgname{Tohoku University}, \postcode{980-8577}, \city{Sendai}, \country{Japan}}

\affil[7]{\orgname{Max-Planck-Institut für Intelligente Systeme}, \postcode{705 69}, \city{Stuttgart}, \country{Germany}}

\affil[8]{\orgname{Helmholtz-Zentrum Berlin für Materialien und Energie}, \postcode{12489}, \city{Berlin}, \country{Germany}}

\abstract{Constriction-based spin-Hall nano-oscillators (SHNOs) show great promise for application as highly tunable microwave sources with straightforward scalability toward large coupled networks. However, details of the magnetization dynamics within SHNOs have thus far not been addressed experimentally, due to the minute time and length scales involved. In this work, we present direct imaging of the magnetization dynamics within a single CoFeB-based SHNO using time-resolved scanning transmission X-ray microscopy. Our measurements reveal that the magnon amplitude is strongest at the two constriction edges, with a pronounced asymmetry favoring one edge, and that the emitted spin waves exhibit strongly anisotropic propagation. Micromagnetic simulations suggest that grain boundaries and the Dzyaloshinskii–Moriya interaction (DMI) play a key role in both effects. Furthermore, the magnetodynamics changed during measurement, indicating that the CoFeB/MgO interface may be more susceptible to X-ray induced modifications than previously recognized, challenging its presumed radiation hardness.}


\maketitle

\section*{Introduction} %Main
The generation and manipulation of collective spin excitations is at heart of spintronics and magnonics, owing to their potential to enable low-consumption processing of digital and analog information~\cite{chumak2014ntcom, chumak2022-advances-spin-wave-computing}, and next-generation low-power non-Von Neumann computational architectures~\cite{finocchio2024roadmap, Gonzalez2024-spintronic-devices-next-generation-computation}. These collective spin excitations, or spin waves (SWs), are able to propagate (and carry information) without Joule heating. Building on this low-consumption propagation, SWs have been explored for miniaturized delay lines~\cite{Bankowski2015-magnonic-delay-line}, boolean logic~\cite{Magnonics_Spin_Wave_Logic_Gates, khitun2011non, alexander2010magnonic}, signal processing~\cite{Papp2016-sw-signal-processing} and wave-based unconventional computing~\cite{Papp2021-NNs-SW-interference, Gartside2022-artificial-spin-ice, Korber2023-pattern-recognition-magnon-reservior, Litvinenko2023-SWIM, Gonzalez2024-global-biasing-SWIM}. 

Nanoconstriction-based spin Hall nano-oscillators (SHNOs) are one of the most promising and versatile devices for SW generation given their simple and mature fabrication process~\cite{kumar2022fabrication}, CMOS compatibility~\cite{Zahedinejad2018apl}, and voltage tunability~\cite{fulara2020natcomm,zahedinejad2022memristive,kumar2022fabrication,Gonzalez2022}. They support multiple dynamical modes, both localized~\cite{mazraati2018pra} and propagating ~\cite{Fulara2019SciAdv, behera2022energy}, and convert a direct current (DC) input into a tunable radio-frequency (RF) output, making them ideal for exploring diverse spin-wave phenomena within a single device. This versatility has also enabled the construction of SHNO arrays susceptible to mutual synchronization, which not only provides a model platform for studying complex dynamics in coupled nonlinear systems \cite{Wittrock2024}, but also improves linewidth and output power, facilitating tunable RF applications~\cite{kumar2023robust}. Furthermore, SHNO arrays might also serve as a potential platform for low-power highly scalable neuromorphic spin-wave computing~\cite{Awad2016natphys,Zahedinejad2020natnano,zahedinejad2022memristive,kumar2023robust}.

A detailed understanding of the magnetization dynamics within spin-Hall nano-oscillators requires experimental techniques capable of resolving nanometer spatial scales and sub-nanosecond temporal dynamics. Conventional approaches—such as Brillouin light scattering (BLS), electrical detection via anisotropic magnetoresistance (AMR), or micromagnetic simulations—have each provided valuable insights into SHNO behavior. However, BLS is limited in spatial resolution, AMR offers only indirect access to the dynamics, and simulations, while powerful, cannot fully replace experimental validation.

In contrast, time-resolved scanning transmission X-ray microscopy (TR-STXM) enables direct visualization of nanoscale magnetization dynamics, capturing both spatial profiles and temporal evolution of spin-wave modes. Even static STXM has revealed SW features not anticipated by theory, e.g. the unexpectedly large size of magnetic droplet solitons and their controlled freezing and thawing~\cite{chung2018direct, Ahlberg2022}. Therefore, extending STXM to the time domain provides an essential tool for understanding the complex behavior of individual oscillators, which in turn governs their coupling in arrays. For instance, recent work on spin-wave-mediated SHNO synchronization has shown that the wavelength and propagation direction of spin waves critically influence the phase relationships between neighboring oscillators~\cite{Kumar2025-spin-wave-mediated-shno-sync}. Direct observation of these dynamics therefore provides the missing link needed to accurately model, design, and optimize coupled SHNO systems.

In this work, we directly observe spin wave auto-oscillation (AO) modes within a constriction-based SHNO using time-resolved STXM. This observation allows us to reevaluate many assumptions of previous studies and shed new light on the emission mechanisms of SWs in SHNOs.

% Moreover, TR-STXM has revealed unexpected features in the dynamic modes of magnetic droplet solitons, demonstrating how direct observation can uncover phenomena that are inaccessible with conventional, time-averaged techniques~\cite{jiang2024magnetic}. \textbf{MA: There is not even static STXM in that paper!}

%SHNOs are usually studied indirectly via electrical anisotropic magnetoresistance, in time-integrated experiments, such as Brillouin light scattering; or by numerical micromagnetic simulations. These experimental techniques lack the necessary spatial and temporal resolution to directly observe the dynamics within an SHNO. In best case scenario, they allow one to make assumptions about the oscillating magnetic thin films. Micromagnetic simulations are then carried out based on these assumptions, and the time-resolved results of those are used to study the magnetization dynamics. Due to the aforementioned resolution limitations, these simulations only focus on the strongest magnetic interactions.

%As shown in a very recent study on spin-wave mediated SHNO synchronization~\cite{Kumar2025-spin-wave-mediated-shno-sync}, the wavelength and propagation direction of SWs drastically affect the phase relation between neighboring oscillators in an array. Therefore, direct observation of the magnetization dynamics of an oscillator is a key to fully understanding their intricate physics. 

\section*{Results and discussion}
\subsection*{Direct observation of auto-oscillation modes}
\begin{figure}[ht]
    \centering
    \includegraphics[width=\textwidth]{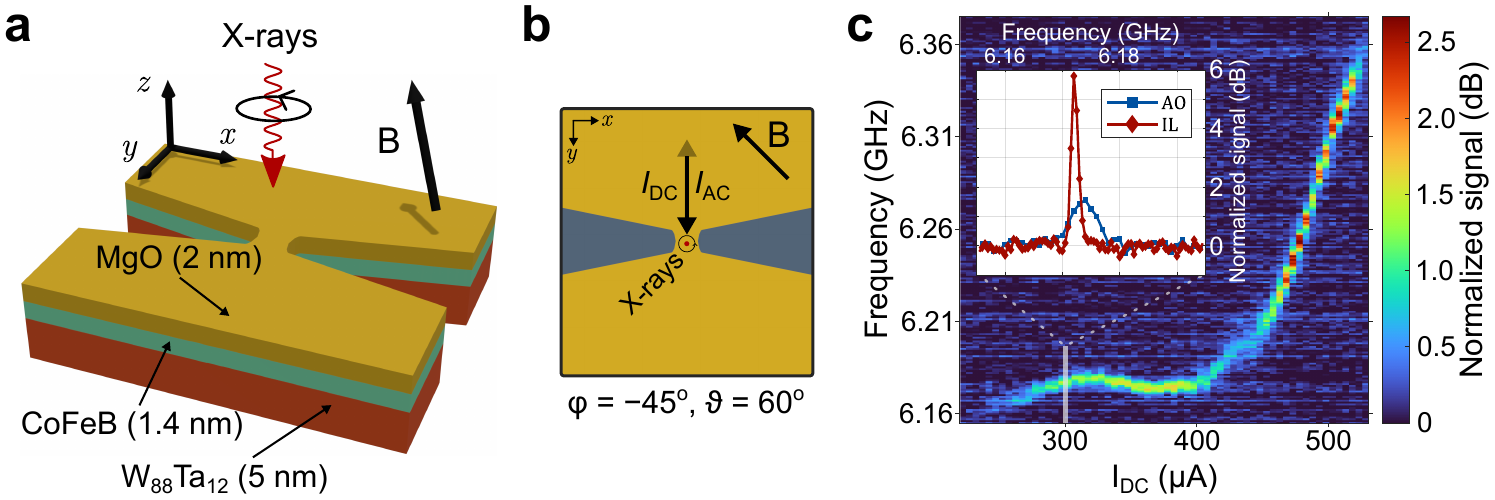}
    \caption{\textbf{Sample and electrical measurement.} \textbf{a} 3D schematic of the SHNO device showing the orientations of the external field and incident X-rays. The external field \textbf{B} had a strength of \SI{255}{\milli\tesla}, in-plane angle $\varphi=\SI{-45}{\degree}$, and out-of-plane angle $\vartheta=\SI{60}{\degree}$. The X-ray incident angle was normal to the sample plane. \textbf{b} Sample top view.  \textbf{c} Power spectral density (PSD) of the AOs as a function of $I_\mathrm{DC}$. The inset shows a cut along the white line of the free running (red diamonds) and injection locked (blue squares) signal at \SI{300}{\micro\ampere}.}
    \label{fig:setup-and-electrical-measurements}
\end{figure}

The samples, constriction-based SHNOs, were fabricated from W$_{88}$Ta$_{12}$(\SI{5}{nm})/Co$_{20}$Fe$_{60}$B$_{20}$(\SI{1.4}{nm})/MgO(\SI{2}{nm}) trilayer stacks and the device design is illustrated in Fig.~\ref{fig:setup-and-electrical-measurements}a--b. A direct current ($I_\mathrm{DC}$) running through the W-Ta heavy metal (HM) generates a spin-orbit-torque (SOT) in the CoFeB layer via the spin Hall effect. In the constriction region, the SOT is large enough to compensate damping and thereby generate a localized magnetization precession called auto-oscillation~\cite{Demidov2014apl}. The MgO layer is used to induce perpendicular magnetic anisotropy (PMA), which is necessary to excite propagating spin waves~\cite{Fulara2019SciAdv, fulara2020natcomm}. A detailed description of the fabrication process is provided in the~\nameref{section:methods} section.

Figure \ref{fig:setup-and-electrical-measurements}c shows the power spectral density (PSD) of the SHNO auto-oscillation as a function of $I_\mathrm{DC}$ in an external field of \SI{255}{\milli\tesla}, applied at oblique in-plane (IP) ($\varphi=\SI{-45}{\degree}$) and out-of-plane (OOP) ($\vartheta=\SI{60}{\degree}$) angles. The onset of AO is observed at approximately \SI{220}{\micro\ampere}, showing a non-monotonic current dependence of the AO frequency, in the range between \SI{220}{\micro\ampere} and \SI{460}{\micro\ampere}. 

This non-monotonic behavior is a characteristic feature of SHNOs with PMA and has been associated with AOs localized close to the constriction due to a negative nonlinearity coefficient, which describes the magnitude and sign of magnon-magnon interactions in the system~\cite{Dvornik2018prappl,Fulara2019SciAdv}. At higher currents, the AO frequency ($f_\mathrm{AO}$) undergoes a blue shift. This is evidence of a modified confinement of the SW mode. As more SOT is exerted (larger $I_\mathrm{DC}$), the magnetization precesses at a larger angle, reducing the demagnetizing field and changing the sign of the magnon-magnon nonlinearity. This change in sign promotes propagation of the spin waves outside the nanoconstriction region and a quasi-linear increase in AO frequency~\cite{Fulara2019SciAdv}. 

The SHNO dynamics can be further stabilized using a process known as second harmonic injection locking. In this process, an external signal with twice the frequency of the AO mode is injected. This signal phase-locks and drives the oscillator at ($f_\mathrm{AO}$), greatly increasing its power and reducing its linewidth~\cite{rajabali2023injection}, as shown in the inset of Fig.~\ref{fig:setup-and-electrical-measurements}c. By injection-locking the device, we are not only drastically increasing the signal strength that can be detected, but also enabling TR-STXM. The operation frequency of the microscope can be synchronized with the AO frequency, allowing time-resolved detection of the magnetization dynamics. The details of the electrical and TR-STXM setups are given in the~\nameref{section:methods} section.

\begin{figure*}
    \includegraphics[width=\textwidth]{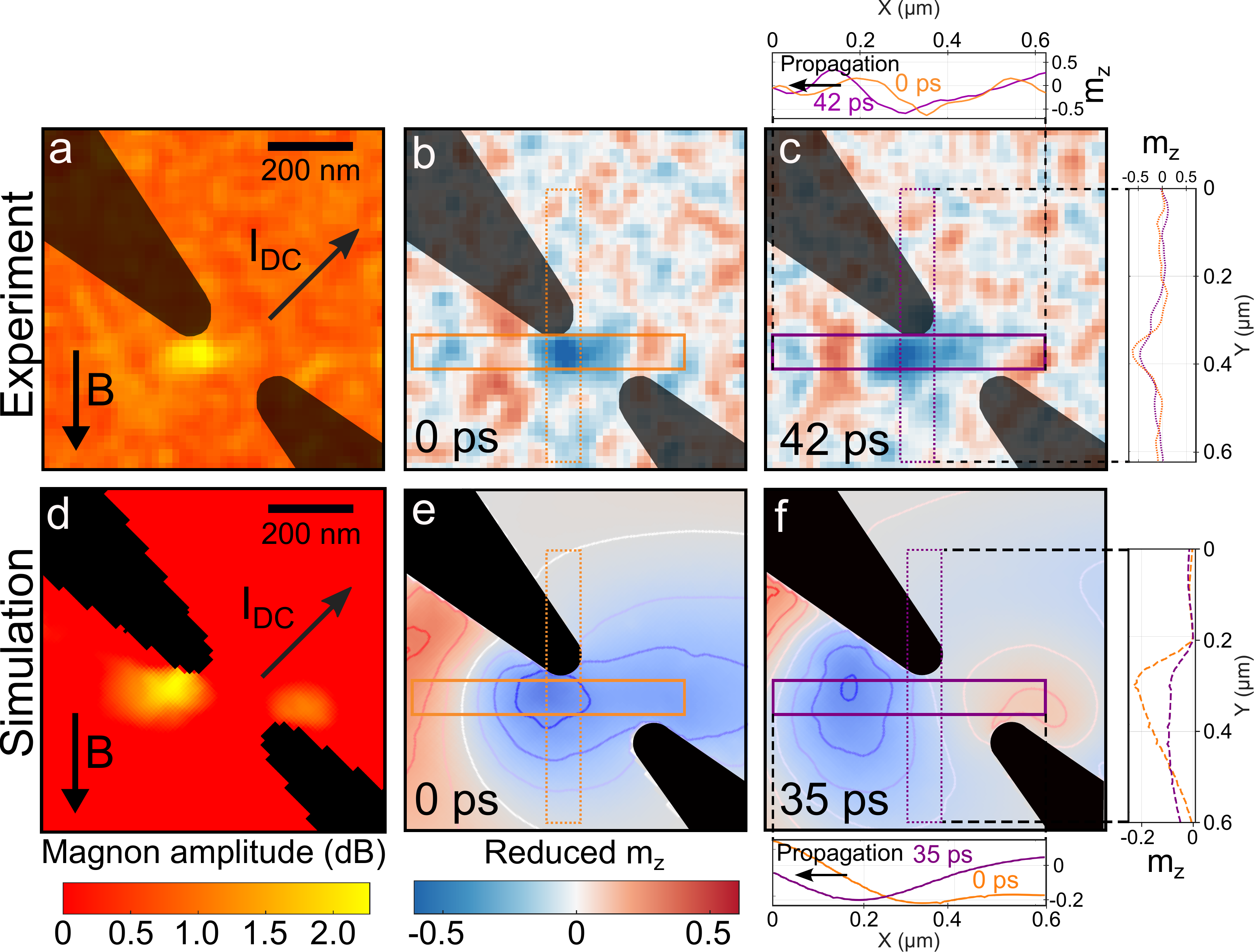}
    \caption{\textbf{Experimental and simulated ps-fast magnetization dynamics} \textbf{a} Measured relative magnon amplitude. The highly asymmetrical amplitude suggests the presence of additional magnetic interactions, not considered by the existing literature. \textbf{b-c} Snapshots of the spin wave propagation. Average magnetization ($m_{\mathrm{z}}$) of the marked rectangular areas is shown in the insets to the right and top. We observe a spin wave propagating to the left in the horizontal direction. \textbf{d-f} Simulated magnon amplitude and snapshots. Agreement between the experiment and the simulation was only achieved with the inclusion of grain boundaries and DMI in the CoFeB layer.}
    \label{fig:X-ray-measurements-and-sims}
\end{figure*}

The TR-STXM measurement was performed at a nominal temperature of \SI{50}{\kelvin} and a \SI{300}{\micro\ampere} driving current. The applied field was kept at the same strength and orientation (\textbf{B} $= \SI{255}{\milli\tesla}$, $\varphi=\SI{-45}{\degree}$, $\vartheta=\SI{60}{\degree}$).  The locking frequency was chosen to be $f_{\mathrm{IL}}=\SI{6.17}{\giga\hertz}$, in accordance with the previous electrical characterization shown in Fig.~\ref{fig:setup-and-electrical-measurements}c. The dynamic OOP magnetization $m_\mathrm{z}$ was measured over an $800 \times 800~\SI{}{\nano\meter\squared}$ square-shaped area around the nanoconstriction. The acquired 31 frames were processed using an FFT-based algorithm to obtain frequency-specific space-resolved magnon amplitude and phase of the signal, which can be looked at separately or recombined into snapshots. Figure~\ref{fig:X-ray-measurements-and-sims}a shows the amplitude map at the AO frequency. We observe the presence of a SW mode localized at the edges of the constriction. These edge modes are characteristic of SHNOs with~\cite{Fulara2019SciAdv} and without~\cite{Dvornik2018prappl} PMA, but our observation shows mode asymmetry that has not been previously reported. %When we include phase information, long-held assumptions about SHNOs must be reevaluated to account for these results.

% Fig3 : Show STXM Measurement with cross section

The phase information is included in the snapshots in Fig.~\ref{fig:X-ray-measurements-and-sims}b--c showing two different points in time. The insets to the top and the side of Fig.~\ref{fig:X-ray-measurements-and-sims}c display the average magnetization in the areas of the horizontal and vertical rectangles, respectively. A directed SW propagation perpendicular to the applied field can be observed, with the wavefront moving horizontally to the left (upper inset). Parallel to \textbf{B}, the wave remains in the same position (side inset). An animation of the time-resolved data is provided in the supplemental material.  

The anisotropic propagation is surprising and highlights previously undetected features of the local magnon potential landscape in SHNOs. The observation has significant implications, as it challenges existing assumptions and may prompt a reevaluation of our fundamental understanding of the intrinsic dynamics of a single oscillator, ultimately affecting how we model, design, and optimize SHNO-based devices.

%This SW propagation anisotropy in SHNOs has not been reported before either experimentally or in numerical simulations. Our method has allowed us to reveal previously unknown details of the potential landscape the propagating magnons are subjected to in SHNOs. 

%\subsection*{Switching micromagnetic parameters}
\subsection*{Anisotropic spin wave propagation}

The direct imaging of the spin wave modes reveals unexpected localization and propagation asymmetries. To better understand these features, we modeled the measured SW dynamics using micromagnetic simulations, as demonstrated in Fig.~\ref{fig:X-ray-measurements-and-sims}d--f. All parameters of the model are provided in the~\nameref{section:methods} section. In order to reproduce the observed magnetization dynamics, we had to lower the measured (blanket film) PMA, add grain boundaries, and include interfacial Dzyaloshinskii-Moriya interaction (DMI) in the magnetodynamical layer. To better understand the role of each of these material parameters, we performed simulations adding each subsequent component, and studied the role each plays on the shape of the SW modes.

\begin{figure*}[ht]
    \centering
    \includegraphics[width=\linewidth]{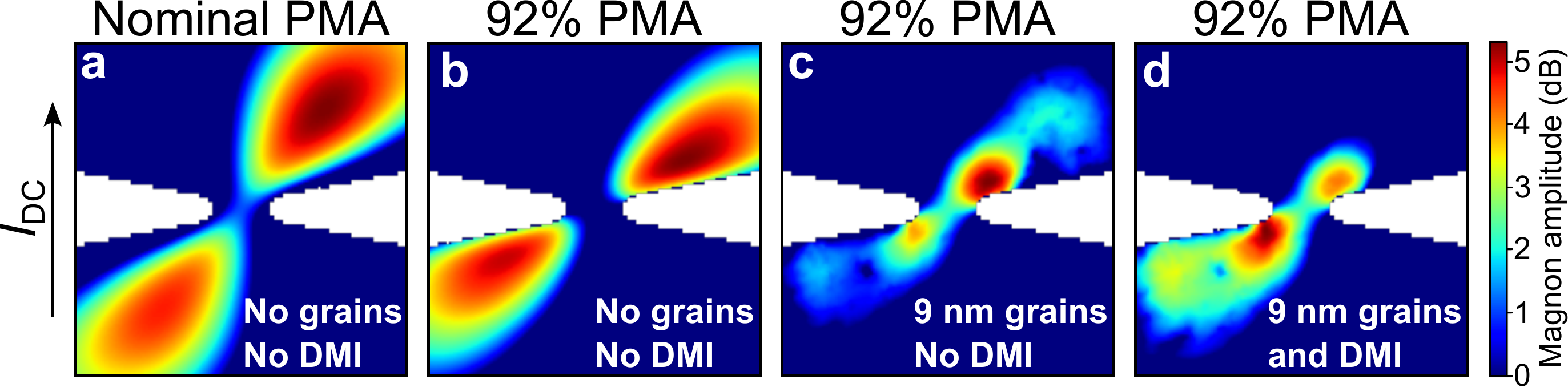}
    \caption{\textbf{AO modes for different simulation parameters.} \textbf{a} Nominal PMA and monocrystalline film, similar to the existing literature. \textbf{b} Reduced PMA. \textbf{c} Reduced PMA with grain boundaries. \textbf{d} Reduced PMA with grain boundaries and DMI. In order to reproduce the asymmetry and SW direction detected in the measurements, grain boundaries and DMI must be considered. $I_\mathrm{DC}$ flows in the direction indicated by the arrow on the left.}
    \label{fig:component-toggling}
\end{figure*}

Figure~\ref{fig:component-toggling}a shows the simulated magnon amplitudes for a CoFeB-layer based on the existing literature~\cite{dvornik2011numerical,Fulara2019SciAdv,fulara2020natcomm} and the experimental PMA value (blanket film, before fabrication). Although there is a slight amplitude asymmetry, the shape of the modes is symmetrical and extended, as expected for propagating SW edge modes. In contrast, the modes observed in the synchrotron measurement are more localized, just off the side of the constriction, and have a more pronounced IP propagation anisotropy (Fig.~\ref{fig:X-ray-measurements-and-sims}).

By refining the PMA value in our simulations, the modes start to resemble the measurement. Weaker PMA in devices compared to the as-grown film has been observed before in SHNOs, and the effect is more pronounced for smaller nanoconstrictions~\cite{Fulara2019SciAdv}.  We suspect that the fabrication procedure, specifically the Ar ion milling, affects the CoFeB/MgO interface and thereby reduces the PMA. It has been reported that ion irradiation can introduce defects at CoFeB/MgO interfaces and reduce the PMA~\cite{Coi2021-heavy-ion-irradiation-PMA, Juge2021-He-ions-skyrmions, Kern2022-he-ion-generation-and-guiding-skyrmions}, which is known to be sensitive to random rearrangements of the crystalline lattice~\cite{KONOBEYEV2017-PMA-suscetible-to-lattice-displacements}. Other possible mechanisms are Ta diffusion into the CoFeB~\cite{Chatterjee2019-physicochemical-origins-improvement-STT-RAM} and oxidation of the magnetodynamical layer at the patterned edges~\cite{kinoshita2014-process-induced-PMA-reduction}. Regardless of the PMA reduction's origin, the SW modes are very susceptible to the local magnetic environment, and modifications of the anisotropy landscape certainly modifies the localization and propagation direction of the AO modes, as seen in Fig.~\ref{fig:X-ray-measurements-and-sims}a--b.  In the simulations, an 8\% decrease in PMA delocalizes the oscillating modes from the constriction region, which displaces the origin and propagation direction of the SWs. The specific percentage number was found by performing micromagnetic simulations with different PMA values and matching the AO mode profiles with the experiment.

Previous simulations on realistic polycrystalline constriction-based SHNOs found that the presence of grain boundaries indeed changes the AO modes~\cite{Capriata2022-grain-boundaries-SHNOs}. In this work, the presence of grain boundaries lowers the current threshold for the onset of AOs. A simulated monocrystalline device requires \SI{350}{\micro\ampere} to autooscillate, in contrast to the less than \SI{300}{\micro\ampere} necessary both experimentally and in our final simulations. More importantly, the shortened propagation length due to magnon scattering at the grain boundaries localizes the edge modes and sharply increases the difference in relative power between the two edges. In the measurement, the low amplitude mode can easily be missed due to the low signal-to-noise ratio. However, it is present upon careful observation of Fig.~\ref{fig:X-ray-measurements-and-sims}b--c. Negative $m_{\mathrm{z}}$ SWs (shown in blue) propagate to the left, while the weaker positive $m_{\mathrm{z}}$ waves (shown in red) propagate in the opposite direction as seen in the right end of the horizontal rectangles. Figure~\ref{fig:X-ray-measurements-and-sims}a also displays a faint structure at the predicted position. Our average grain diameter of \SI{9}{\nano\meter} (found by simulating different grain sizes) is consistent with the previously reported experimental value for similar Co$_{60}$Fe$_{20}$B$_{20}$ thin films~\cite{CHEN2010-effect-grain-size-CoFeB-films}.  

It is well established that heavy metal-ferromagnet heterostructures leads to Dzyaloshinskii-Moriya interaction ($D$)~\cite{Cheng2019prb, Jaiswal2017apl, Tacchi2017prl, Torrejon2014-DMI-CoFeB-MgO}. Yet, DMI is most often neglected in experimental and numerical treatments of nanoconstriction SHNOs. In our simulations, we used an experimental value of $D=$~\SI{250}{\micro\joule\per{\meter\squared}} reported for ultrathin W/CoFeB/MgO heterostructures~\cite{Torrejon2014-DMI-CoFeB-MgO}, and the results show an excellent match to the measurement (Fig.~\ref{fig:X-ray-measurements-and-sims}). When DMI is included in the model, the SW propagation direction is reversed, and the asymmetry of the edge modes becomes more pronounced. To be discerned, these features require the detailed information obtained from direct imaging. This explains why, although DMI has been mentioned in previous SHNO studies~\cite{Ranjbar2014-CoFeB-SHNOs, fulara2020natcomm}, this is the first time it must be explicitly included in order to reproduce the experimental results. Consequently, TR-STXM has allowed us to develop a more nuanced understanding of the operation of real-world devices in a way that previous indirect or time-integrated measurements could not. In turn, we expect that these results can guide SHNO design and fabrication for microwave and neuromorphic applications.

\begin{figure}
    \centering
    \includegraphics[width=0.5\linewidth]{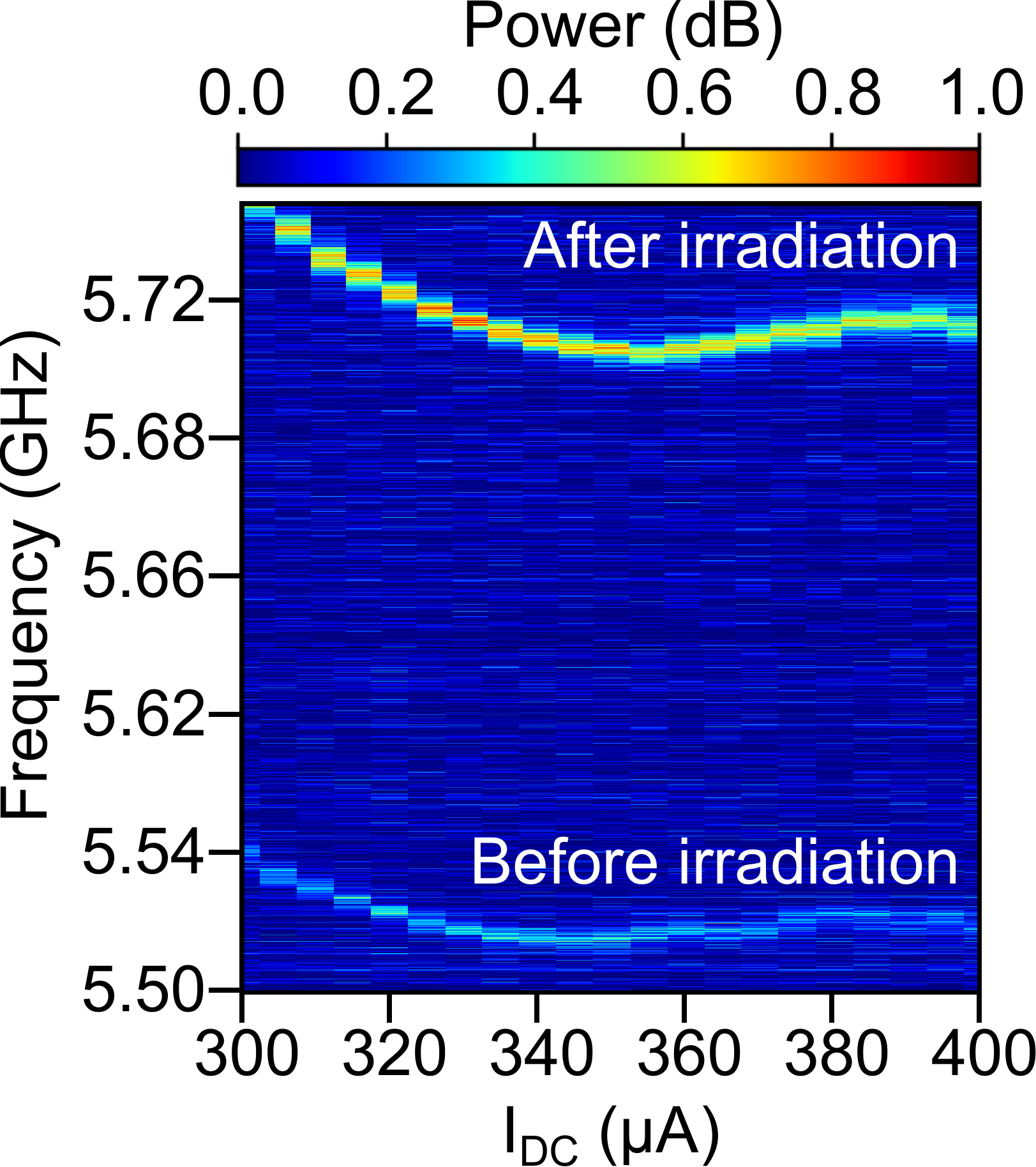}
    \caption{\textbf{AO modes before and after X-ray irradiation.} Power spectral density as function of current. The frequency and amplitude increase after irradiation, hinting at larger values of the PMA and SOT. The modified curvature implies a transformation of the mode geometry. While a rising frequency was consistently observed, the changes in amplitude and curvature varied between samples.}
    \label{fig:irradiation-modes}
\end{figure}

\subsection*{Irreversible X-ray effects on CoFeB magnetodynamics}
Exposure to X-rays irreversibly alters the CoFeB layer, making a systematic study of the magnetodynamics challenging. Several samples with different composition and thicknesses of the CoFeB layer have been studied, and in each case it was found that the AO curve changes significantly upon X-ray illumination. Both the shape as well as the general frequency range changed, which was often accompanied by a decrease in AO amplitude. The signal strength briefly increased before decreasing steadily, and suddenly dropping.

Figure~\ref{fig:irradiation-modes} depicts an example of X-ray induced changes in the magnetodynamics. The AO frequency rises by hundreds of \SI{}{\mega\hertz}, and both the amplitude and curvature increase. However, the increase of the latter two is not consistent across all samples, instead they often decrease when irradiated. 
The acquisition of TR images (Fig.~\ref{fig:X-ray-measurements-and-sims}) was performed before major degradation occured, ensuring the reliability of the observed AO modes. 
Long-term imaging, however, was limited because prolonged exposure alters device frequencies, causing a loss of injection-locking and microscope synchronization. 
Importantly, these observations also provide an indirect probe of the radiation hardness of CoFeB/MgO-based SHNOs, revealing how extended X-ray exposure affects their magnetic properties. The possible origins and implications of this behavior are discussed below. 

%Note that the sharp degradation process, after a small irradiation threshold, makes TR-STXM unreliable, since the injection-locking vanishes and the microscope loses synchronization to the AOs. Measurements are only possible if the process is sufficiently slow. The possible origin and consequences of the degradation are diverse, and we expand upon it below.

Naturally, the simplest explanation of the X-ray induced alterations of the magnetic properties is radiation-induced heating. To put this at test, we reduced the photon flux on the sample. This allows any 
temperature gradient to dissipate before it can induce further thermal annealing in the CoFeB. However, the PMA-changes solely depended on the dose (time-integrated flux). Smaller doses and longer radiation intervals produce consistent degradation in all samples. Furthermore, the effect occurs sharply, while thermal annealing is a gradual process that requires a sample to be kept at high temperatures for a longer period of time. Thus, the observations of briefly modified properties followed by an irreversible degradation are only consistent with chemical changes induced by the X-rays.

The AO frequency of SNHOs is closely related to the ferromagnetic resonance (FMR)~\cite{Jiang2024review}, which is largely governed by the external field, saturation magnetization and anisotropy~\cite{Kittel1948, Farle1998}. The anisotropy in our samples is dominated by the PMA induced at the CoFeB/MgO interface. This effect is achieved via symmetry breaking and hybridization of the Fe(Co)-3d and O-2p orbitals at the metal/oxide interface~\cite{Niranjan2010-electric-field-effect-Fe-MgO, Yang2011-first-principles-PMA,Khoo2013-first-principles-PMA,Leutenantsmeyer2015}. In order to hybridize, the Fe and Co atoms should be in the crystalline Co-Fe form, rather than the amorphous CoFeB. During the annealing process at fabrication, B diffuses into the HM metal layer, with Ta acting as a boron sink, which allows Co-Fe to form. The value of the PMA will depend on these two factors: the crystallinity at the metal/oxide interface and the B diffusion into the HM layer. By understanding the effect of the X-rays on both, we can explain the irreversible changes induced during our measurements.

In our case, the electromagnetic radiation most likely modifies the metal/oxide interface. High doses of X-rays can generate oxygen vacancies in the oxide at CoFeB/MgO~\cite{He2022-high-X-ray-dose-MgO-degradation}. The same ionizing process could also reduce Co and Fe oxides at the interface~\cite{Cheng2024-soft-X-ray-damage}, allowing O to migrate back into the MgO, enhancing the crystallinity and thereby the PMA, leading to the effect shown in Fig.~\ref{fig:irradiation-modes} (higher $f_{\mathrm{AO}}$ and amplitude). It is worth mentioning that NiFe (another material commonly used in SHNOs) does not show such degradation and in this respect, exhibits a better resilience to radiation.

Another method to modify interfaces is ion bombardment. Devolder~et~al. studied the effect on Ta/CoFeB/MgO stacks, showing that small fluences mitigate B diffusion and enhance the PMA, likely due to CoFeB crystallization~\cite{Devolder2013-irradiation-induced-PMA-CoFeB-MgO}. A similar process could be taking place in our samples, with the X-rays generating both oxygen ions and photoelectrons that drive B diffusion. In larger doses, the radiation would cause a breakdown of the MgO due to oxygen vacancies, which is also consistent with our experiments. 

Our observations challenge the often-promoted claim of radiation hardness of CoFeB/MgO-based spintronic devices (i.e., that X-ray or ionizing radiation does not modify their magnetic properties)~\cite{Electromagnetic_Radiation_On_MTJs}. 
Radiation hardness is particularly relevant for technological applications of spintronic devices in high-radiation environments, such as spacecraft, satellites, or particle accelerators, where exposure to ionizing radiation could unintentionally alter magnetic properties and compromise device performance. 
Additional studies are needed to fully clarify the mechanism of the X-ray-induced annealing, which is of considerable interest given the scientific and technological significance of CoFeB/MgO heterostructures~\cite{Shao2021ieeetmag, STT-MRAM2024}. SHNOs might be an ideal system for further studies, since the amplitude and frequency of their AOs are sensitive to small changes in the local effective magnetic field~\cite{fulara2020natcomm, Gonzalez2022, Kumar2025-spin-wave-mediated-shno-sync} and can be detected using electrical or time-integrated optical techniques. 

%MA Our observations and numerical simulations challege the assumption of the radiation hardness (i.e. the claim that X-ray radiation does not modify the magnetic material) of CoFeB/MgO interfaces. Further studies using X-ray experiments have the potential to shed light on the mechanism of the X-ray-induced annealing given the generic nature and technological importance of the CoFeB/MgO not only for SHNOs but also for magnetic tunneling junctions, a very mature type of spintronic oscillator. 

%MA A more comprehensive study is necessary to fully understand the role of the X-ray dose on this material stack and how to improve radiation hardness in these multiphysical stacks. SHNOs might be ideal systems for studying the effects of this X-ray annealing on the multilayer stack since the amplitude and frequency of AOs are sensitive to small changes in the components of the local effective magnetic field~\cite{fulara2020natcomm, Gonzalez2022, Kumar2025-spin-wave-mediated-shno-sync} and can be detected using electrical or time-integrated optical techniques. 

%\section*{Discussion}
\vspace{\baselineskip}

\noindent In summary, SHNOs are central in the advancement of scalable spintronic microwave devices. By using TR-STXM we have directly observed AO modes in an SHNO and recorded their time evolution with exceptional spatial resolution. The measurements were reproduced by micromagnetic simulations, which demonstrated that commonly overlooked parameters are, in fact, important. A fabrication-induced reduction of PMA together with polycrystallinity (grains) displaces the AO modes and causes amplitude asymmetry between them, whereas DMI yields anisotropic SW propagation. In addition, we found that X-ray radiation irreversibly and constantly modifies the effective anisotropy, making more in-depth characterizations of CoFeB/MgO stacks challenging. Overall, the results show the need to revise the physical models underlaying SHNOs and their mutual synchronization. These insights contribute to a better understanding of the individual and collective oscillation dynamics, with implications for accelerate the design of spintronic complex networks for applied next-generation computing architectures.

%As an example, recent research shows that the synchronization mechanism in SHNO arrays is deeply linked with the shape of AO modes and the propagation of SW between oscillators.~\cite{Kumar2025-spin-wave-mediated-shno-sync}

\section*{Methods}
\label{section:methods}
% Fig1 : Show electrical (coarse) setup and sample design
% Sample Used : S347 - 221025 -  NB01 - SiN/Al2O3(3)/W88Ta12(5)/CoFeB(1.4)/MgO(2)/Al2O3(4)

\subsection*{Sample Preparation}
AlO$_{x}$(\SI{3}{nm})/W$_{88}$Ta$_{12}$(\SI{5}{nm})/Co$_{20}$Fe$_{60}$B$_{20}$(\SI{1.4}{nm})/MgO(\SI{2}{nm}) stacks were grown on SiN coated p-Si(100) substrates containing membrane slits of size $20 \times 300~\SI{}{\micro\meter\squared}$ and thickness \SI{200}{\nano\meter}. The growth environment of all layers in the stacks were similar to previously reported works~\cite{behera2022energy, behera2024,Kumar2025-spin-wave-mediated-shno-sync}. Subsequently, the samples were annealed at \SI{300}{\celsius} for 1~hour to crystallize the CoFeB and MgO at the interface. After growth, the stacks were patterned into SHNOs of three different nano-constriction widths, \SI{120}{nm}, \SI{150}{nm} and \SI{200}{nm} using E-beam lithography (EBL) with the negative electron resist, hydrogen silsesquioxane (HSQ, Product number: XR-1541-002 (2\%)). Ar-ion beam tool from Oxford instruments (Ionfab 300 Plus) was used for etching the negative patterns. Next, the negative resist was removed, and ground-signal-ground (GSG) co-planar wave-guide (CPW) masks were defined by an optical lift-off lithography process. Finally, the CPW was deposited using a sputtered Cu(\SI{800}{\nano\meter})/Pt(\SI{20}{\nano\meter}) bilayer. It is worth reiterating that due to the radiation induced irreversible changes mentioned in the manuscript, only the \SI{200}{nm} sample was measured successfully using STXM.

\subsection*{Electrical Setup}
For electrical characterization and injection-locking of the device, the sample was connected to the electrical setup depicted in Fig.~\ref{fig:setup-and-electrical-measurements}a. A sourcemeter (Keithley 2450 SMU) is used to supply a direct current $I_\mathrm{DC}$ to induce AOs in the nanoconstriction. The AO then induces an AC signal through the anisotropic magnetoresistance (AMR) effect. This signal passes the bias tee and a circulator, is amplified using a low noise amplifier (LNA) with \SI{72}{\decibel} gain, and finally detected using a spectrum analyzer (Agilent E4407B). To achieve injection locking, an arbitrary waveform generator (AWG) (Keysight M8195A) combined with a YIG bandpass filter and an amplifier were used to generate an AC signal with the desired locking frequency, which goes through the circulator and bias tee to reach the sample. IL is realized by twice the AO frequency (2f-locking), which makes it possible to detect the increased AMR signal at 1f in the spectrum analyzer.\\

\subsection*{TR-STXM Setup}
The electrical setup was connected to the sample via a high frequency sample holder, which itself was positioned inside the vacuum chamber of the X-ray microscope {\tt{MAXYMUS}} at the {\tt{BESSYII}} electron storage ring operated by the {\tt{Helmholtz-Zentrum Berlin für Materialien und Energie}}~\cite{Weigand2022-maxymus}. For TR-STXM measurements, the sample was illuminated with monochromatic x-rays from an undulator, with an energy corresponding to the Fe L$_3$ edge, making it possible to exploit the X-ray magnetic circular dichroism (XMCD) effect to obtain magnetic contrast. Furthermore, the time structure of the synchrotron was utilized in an asynchronous pump-probe scheme to achieve temporal resolution. This was used to directly observe magnetization dynamics on time and length scales required for nanoconstriction based SHNOs. A set of four rotatable permanent magnets allowed for controlling the magnetic field at the sample position, while a liquid helium-flow cryostat enabled temperature control down to \SI{20}{\kelvin}.\\

\subsection*{Post-processing of TR-STXM Measurements}
The TR-STXM measurements yield snapshots of the magnetization landscape at equidistant time steps, in this case 31 frames per dynamic cycle. To significantly increase the signal to noise ratio, these 31 frames were used to perform a fast Fourier transform (FFT), which transforms the signal into the frequency domain. The fundamental frequency was selected to eliminate noise at different frequencies, and a phase-resolved snapshot was created by using the amplitude and phase of the Fourier transform. The dynamic representation of the data (video in the supplemental material) was retrieved by plotting the FFT data at different time steps. The evaluation of the data was similar to the procedure described in Ref.~\cite{gross2021miep}. 

\subsection*{Micromagnetic simulations}
Micromagnetic simulations have been carried out using the GPU-accelerated \emph{mumax$^3$} solver \cite{mumax3}, with the current density, Oersted and demagnetization field inputs calculated using the multiphysics {\tt{{COMSOL}}} software \cite{COMSOL}. The  CoFeB layer was discretized using a 1024$\times$1024$\times$1 rectangular mesh with \SI{4}{nm}$\times$\SI{4}{nm}$\times$\SI{1.4}{nm} cellsize and current and field geometry according to the experimental setup shown in Fig.~\ref{fig:setup-and-electrical-measurements}a. The spin-orbit torque was modeled by a Slonczewski term using the input current density and a spin-Hall angle of 0.6. The CoFeB layer parameters were chosen as: saturation magnetization $M_\mathrm{S}=\SI{1050}{\kilo\ampere\per\meter}$, exchange stiffness $A_\mathrm{ex}=\SI{19}{\pico\joule\per\meter}$,  Gilbert damping constant $\alpha_0=9\cdot 10^{-3}$, and gyromagnetic ratio $\gamma/2\pi= \SI{29.1}{\giga\hertz\per\tesla}$, in line with previous work on similar bilayers~\cite{Gonzalez2022}. The grain boundaries were simulated using different magnetization regions generated by Voronoi tesellation with an average diameter of \SI{9}{\nano\meter}, their PMA was modeled using a random normal distribution with average 0.92 of the measured PMA value and a width of 5$\%$. The measured PMA was \SI{645}{\kilo\joule\per\cubic\meter}. The interfacial DMI value was \SI{250}{\micro\joule\per{\meter\squared}}, as reported in the literature~\cite{Torrejon2014-DMI-CoFeB-MgO}.

\section*{Supplemental Material}
The supplemental material includes an animated version of the TR-STXM measurement and the corresponding micromagnetic simulation.

\section*{Acknowledgments}
The authors thank the Zentrum Berlin für Materialen und Energie for the allocation of synchrotron radiation beamtime. The authors also acknowledge support from the Horizon 2020 research and innovation programme (ERC Advanced Grant No. 835068 “TOPSPIN”), and the Knut and Alice Wallenberg Foundation (Dnr. KAW 2022.0079 and KAW 2023.0285). S.W. acknowledges financial support from the Helmholtz Young Investigator Group Program (VH-NG-1520).

\section*{Data availability statement}
The data that support the findings of this study are available from the corresponding author upon reasonable request.

\section*{Code availability} The code used in this study is available from the corresponding author upon reasonable request.

\section*{Author contributions} 
V.H.G. and M.A. performed experiments and simulations. F.S., N.B. A.K. and A.F. fabricated the samples and performed experiments. S.W., MW. and S.W. provided guidance for assembling the experimental setup. All authors contributed to the analysis of the results and writing of the manuscript.

\section*{Competing interests} 
The authors declare no competing interests.

\bibliography{bibliography}% Produces the bibliography via BibTeX.

@STRING{ntc = {Nat.\ Commun.}}

@STRING{ntn = {Nat.\ Nanotechnol.}}

@STRING{pra = {Phys.\ Rev.\ Appl.}}

@STRING{prl = {Phys.\ Rev.\ Lett.}}

@STRING{ss = {Surf. Sci.}}

@article{Bankowski2015-magnonic-delay-line,
    author = {Bankowski, Elena and Meitzler, Thomas and Khymyn, Roman S. and Tiberkevich, Vasil S. and Slavin, Andrei N. and Tang, Hong X.},
    title = {Magnonic crystal as a delay line for low-noise auto-oscillators},
    journal = {Appl. Phys. Lett.},
    volume = {107},
    xnumber = {12},
    pages = {122409},
    year = {2015},
    month = {09},
    issn = {0003-6951},
    doi = {10.1063/1.4931758},
    url = {https://doi.org/10.1063/1.4931758},
}

@Article{Papp2021-NNs-SW-interference,
author={Papp, {\'A}d{\'a}m
and Porod, Wolfgang
and Csaba, Gyorgy},
title={Nanoscale neural network using non-linear spin-wave interference},
journal={Nat. Commun.},
year={2021},
month={Nov},
day={05},
volume={12},
xnumber={1},
pages={6422},
issn={2041-1723},
doi={10.1038/s41467-021-26711-z},
url={https://doi.org/10.1038/s41467-021-26711-z}
}

@Article{Korber2023-pattern-recognition-magnon-reservior,
author={K{\"o}rber, Lukas
and Heins, Christopher
and Hula, Tobias
and Kim, Joo-Von
and Thlang, Sonia
and Schultheiss, Helmut
and Fassbender, J{\"u}rgen
and Schultheiss, Katrin},
title={Pattern recognition in reciprocal space with a magnon-scattering reservoir},
journal={Nat. Commun.},
year={2023},
month={Jul},
day={04},
volume={14},
xnumber={1},
pages={3954},
issn={2041-1723},
doi={10.1038/s41467-023-39452-y},
url={https://doi.org/10.1038/s41467-023-39452-y}
}

@Article{Gartside2022-artificial-spin-ice,
author={Gartside, Jack C.
and Stenning, Kilian D.
and Vanstone, Alex
and Holder, Holly H.
and Arroo, Daan M.
and Dion, Troy
and Caravelli, Francesco
and Kurebayashi, Hidekazu
and Branford, Will R.},
title={Reconfigurable training and reservoir computing in an artificial spin-vortex ice via spin-wave fingerprinting},
journal={Nat. Nanotechnol.},
year={2022},
month={May},
day={01},
volume={17},
xnumber={5},
pages={460-469},
issn={1748-3395},
doi={10.1038/s41565-022-01091-7},
url={https://doi.org/10.1038/s41565-022-01091-7}
}

@Article{Weigand2022-maxymus,
AUTHOR = {Weigand, Markus and Wintz, Sebastian and Gräfe, Joachim and Noske, Matthias and Stoll, Hermann and Van Waeyenberge, Bartel and Schütz, Gisela},
TITLE = {TimeMaxyne: A Shot-Noise Limited, Time-Resolved Pump-and-Probe Acquisition System Capable of 50 GHz Frequencies for Synchrotron-Based X-ray Microscopy},
JOURNAL = {Crystals},
VOLUME = {12},
YEAR = {2022},
xnumber = {8},
ARTICLE-xnumber = {1029},
URL = {https://www.mdpi.com/2073-4352/12/8/1029},
ISSN = {2073-4352},
DOI = {10.3390/cryst12081029}
}

@ARTICLE{Shao2021ieeetmag,  
author={Shao, Qiming and Li, Peng and Liu, Luqiao and Yang, Hyunsoo and Fukami, Shunsuke and Razavi, Armin and Wu, Hao and Wang, Kang and Freimuth, Frank and Mokrousov, Yuriy and Stiles, Mark D. and Emori, Satoru and Hoffmann, Axel and {\AA}kerman, Johan and Roy, Kaushik and Wang, Jian-Ping and Yang, See-Hun and Garello, Kevin and Zhang, Wei},  
journal =	{IEEE Trans. Magn.},   
title={Roadmap of Spin–Orbit Torques},   
year={2021},  
volume={57},  
xnumber={7},  
pages={1-39},  
doi={10.1109/TMAG.2021.3078583}
}

@article{kumar2022fabrication,
  title={Fabrication of voltage-gated spin {H}all nano-oscillators},
  author={Kumar, Akash and Rajabali, Mona and Gonz{\'a}lez, Victor Hugo and Zahedinejad, Mohammad and Houshang, Afshin and {\AA}kerman, Johan},
  journal={Nanoscale},
  year={2022},
  volume={14},
  pages={1432--1439},
  url={https://pubs.rsc.org/en/content/articlehtml/2022/nr/d1nr07505e},
  publisher={Royal Society of Chemistry}
}

@article{finocchio2024roadmap,
  title={Roadmap for unconventional computing with nanotechnology},
  author={Finocchio, Giovanni and Incorvia, Jean Anne C and Friedman, Joseph S and Yang, Qu and Giordano, Anna and Grollier, Julie and Yang, Hyunsoo and Ciubotaru, Florin and Chumak, Andrii V and Naeemi, Azad J and others},
  journal={Nano Futures},
  volume={8},
  xnumber={1},
  pages={012001},
  year={2024},
  publisher={IOP Publishing}
}

@INPROCEEDINGS{Papp2016-sw-signal-processing,
  author={Papp, {\'A}dam and Csaba, György and Porod, Wolfgang},
  booktitle={2016 IEEE International Conference on Rebooting Computing (ICRC)}, 
  title={Optically-inspired computing based on spin waves}, 
  year={2016},
  volume={},
  xnumber={},
  pages={1-4},
  doi={10.1109/ICRC.2016.7738707}}

@article{chumak2022-advances-spin-wave-computing,
  title={Advances in magnetics roadmap on spin-wave computing},
  author={Chumak, Andrii V and Kabos, Pavel and Wu, Mingzhong and Abert, Claas and Adelmann, Christoph and Adeyeye, AO and {\AA}kerman, J and Aliev, Farkhad G and Anane, Abdelmadjid and Awad, A and others},
  journal={IEEE Trans. Magn.},
  volume={58},
  xnumber={6},
  pages={1--72},
  year={2022},
  publisher={IEEE}
}

@article{Zahedinejad2020natnano,
	title = {Two-dimensional mutually synchronized spin {Hall} nano-oscillator arrays for neuromorphic computing},
	author = {Zahedinejad, Mohammad and Awad, Ahmad A. and Muralidhar, Shreyas and Khymyn, Roman and Fulara, Himanshu and      Mazraati, Hamid and Dvornik, Mykola and {\AA}kerman, Johan},
        volume = {15},
	issn = {1748-3387, 1748-3395},
	language = {en},
	xnumber = {1},
	urldate = {2020-01-19},
	journal = ntn,
	month = jan,
	year = {2020},
	pages = {47--52},
	doi={10.1038/s41565-019-0593-9}
}

@phdthesis{dvornik2011numerical,
  title={Numerical investigations of spin waves at the nanoscale},
  author={Dvornik, M.},
  year={2011},
  publisher={University of Exeter},
  school={University of Exeter},
  url={https://ore.exeter.ac.uk/repository/handle/10036/3304},
}

@article{Dvornik2018prappl,
  title = {Origin of magnetization auto-oscillations in constriction-based spin {Hall} Nano-Oscillators},
  author = {Dvornik, M. and Awad, A. A. and \AA{}kerman, J.},
  journal = pra,
  volume = {9},
  issue = {1},
  pages = {014017},
  numpages = {9},
  year = {2018},
  month = {Jan},
  publisher = {American Physical Society},
  doi = {10.1103/PhysRevApplied.9.014017},
}

@article{Demidov2014apl,
author = {Demidov, V. E. and Urazhdin, S. and Zholud, A. and Sadovnikov, A. V. and Demokritov, S. O.},
issn = {0003-6951},
journal = {Appl. Phys. Lett.},
xnumber = {17},
pages = {172410},
title={Nanoconstriction-based spin-{Hall} nano-oscillator},
volume = {105},
year = {2014},
doi={10.1063/1.4901027}
}

@article{Awad2016natphys,
author = {Awad, A. A. and D{\"{u}}rrenfeld, P. and Houshang, A. and Dvornik, M. and Iacocca, E. and Dumas, R. K. and {\AA}kerman, J.},
issn = {1745-2473},
journal = {Nat. Phys.},
xnumber = {November},
title={Long-range mutual synchronization of spin {Hall} nano-oscillators},
year = {2017},
pages = {292-299},
volume = {13},
doi={10.1038/nphys3927}
}

@article{Kittel1948,
  title = {On the Theory of Ferromagnetic Resonance Absorption},
  author = {Kittel, C.},
  journal = {Phys. Rev.},
  volume = {73},
  issue = {2},
  pages = {155},
  year = {1948},
  month = {Jan},
  publisher = {American Physical Society},
}

@article{Zahedinejad2018apl,
author = {M. Zahedinejad and H. Mazraati and H. Fulara and J. Yue and S. Jiang and A. A. Awad and J. {\AA}kerman},
title={{CMOS} compatible {W/CoFeB/MgO} spin {Hall} nano-oscillators with wide frequency tunability},
journal = {Appl. Phys. Lett.},
volume = {112},
xnumber = {13},
pages = {132404},
year = {2018},
doi={10.1063/1.5022049}
}

@article{chung2018direct,
  title={Direct observation of Zhang-Li torque expansion of magnetic droplet solitons},
  author={Chung, Sunjae and Le, Q Tuan and Ahlberg, Martina and Awad, Ahmad A and Weigand, Markus and Bykova, Iuliia and Khymyn, Roman and Dvornik, Mykola and Mazraati, Hamid and Houshang, Afshin and others},
  journal=prl,
  volume={120},
  xnumber={21},
  pages={217204},
  year={2018},
  publisher={APS}
}

@article{chumak2014ntcom,
  title={Magnon transistor for all-magnon data processing},
  author={Chumak, Andrii V and Serga, Alexander A and Hillebrands, Burkard},
  journal=ntc,
  volume={5},
  pages={4700},
  year={2014},
  publisher={Nature Publishing Group}
}

@article{mazraati2018pra,
  title={Auto-oscillating spin-wave modes of constriction-based spin Hall nano-oscillators in weak in-plane fields},
  author={Mazraati, Hamid and Etesami, Seyyed Ruhollah and Banuazizi, Seyed Amir Hossein and Chung, Sunjae and Houshang, Afshin and Awad, Ahmad A and Dvornik, Mykola and {\AA}kerman, Johan},
  journal=pra,
  volume={10},
  xnumber={5},
  pages={054017},
  year={2018},
  publisher={APS}
}

@article{Fulara2019SciAdv,
	title = {Spin-orbit torque{\textendash}driven propagating spin waves},
	volume = {5},
	issn = {2375-2548},
	language = {en},
	xnumber = {9},
	urldate = {2019-09-30},
	journal = {Sci. Adv.},
	author = {Fulara, H. and Zahedinejad, M. and Khymyn, R. and Awad, A. A. and Muralidhar, S. and Dvornik, M. and {\r A}kerman, J.},
	month = sep,
	year = {2019},
	pages = {eaax8467},
	URL = {10.1126/sciadv.aax8467}
}

@ARTICLE{Coi2021-heavy-ion-irradiation-PMA,
  author={Coi, Odilia and Pendina, Gregory Di and Sousa, Ricardo and Adrianjohany, Nomena and Dangla, David and Ecoffet, Robert and Torres, Lionel},
  journal={IEEE Trans. Nucl. Sci.}, 
  title={Heavy-Ion Irradiation Effects on Advanced Perpendicular Anisotropy Spin-Transfer Torque Magnetic Tunnel Junction}, 
  year={2021},
  volume={68},
  xnumber={5},
  pages={588-596},
  doi={10.1109/TNS.2021.3071257}}

@article{KONOBEYEV2017-PMA-suscetible-to-lattice-displacements,
title = {Evaluation of effective threshold displacement energies and other data required for the calculation of advanced atomic displacement cross-sections},
journal = {Nucl. Energy Technol.},
volume = {3},
xnumber = {3},
pages = {169-175},
year = {2017},
issn = {2452-3038},
doi = {https://doi.org/10.1016/j.nucet.2017.08.007},
url = {https://www.sciencedirect.com/science/article/pii/S2452303816301224},
author = {A.Yu. Konobeyev and U. Fischer and Yu.A. Korovin and S.P. Simakov}
}

@article{kinoshita2014-process-induced-PMA-reduction,
  title={Process-induced damage and its recovery for a {C}o{F}e{B}--{MgO} magnetic tunnel junction with perpendicular magnetic easy axis},
  author={Kinoshita, Keizo and Honjo, Hiroaki and Fukami, Shunsuke and Sato, Hideo and Mizunuma, Kotaro and Tokutome, Keiichi and Murahata, Michio and Ikeda, Shoji and Miura, Sadahiko and Kasai, Naoki and others},
  journal={Jpn. J. Appl. Phys.},
  volume={53},
  xnumber={10},
  pages={103001},
  year={2014},
  publisher={IOP Publishing}
}

@Article{Torrejon2014-DMI-CoFeB-MgO,
author={Torrejon, Jacob
and Kim, Junyeon
and Sinha, Jaivardhan
and Mitani, Seiji
and Hayashi, Masamitsu
and Yamanouchi, Michihiko
and Ohno, Hideo},
title={Interface control of the magnetic chirality in {CoFeB/MgO} heterostructures with heavy-metal underlayers},
journal={Nat. Commun.},
year={2014},
month={Aug},
day={18},
volume={5},
xnumber={1},
pages={4655},
issn={2041-1723},
doi={10.1038/ncomms5655},
url={https://doi.org/10.1038/ncomms5655}
}

@ARTICLE{Ranjbar2014-CoFeB-SHNOs,

  author={Ranjbar, M. and D{\"u}rrenfeld, P. and Haidar, M. and Iacocca, E. and Balinskiy, M. and Le, T. Q. and Fazlali, M. and Houshang, A. and Awad, A. A. and Dumas, R. K. and {\AA}kerman, J.},

  journal={IEEE Magn. Lett.}, 

  title={{CoFeB}-Based Spin {H}all Nano-Oscillators}, 

  year={2014},

  volume={5},

  xnumber={},

  pages={1-4},

  doi={10.1109/LMAG.2014.2375155}}

@article{Khoo2013-first-principles-PMA,
  title = {First-principles study of perpendicular magnetic anisotropy in {CoFe/MgO} and {CoFe/Mg$_{3}$B$_{2}$O$_{6}$} interfaces},
  author = {Khoo, K. H. and Wu, G. and Jhon, M. H. and Tran, M. and Ernult, F. and Eason, K. and Choi, H. J. and Gan, C. K.},
  journal = {Phys. Rev. B},
  volume = {87},
  issue = {17},
  pages = {174403},
  numpages = {6},
  year = {2013},
  month = {May},
  publisher = {American Physical Society},
  doi = {10.1103/PhysRevB.87.174403},
  url = {https://link.aps.org/doi/10.1103/PhysRevB.87.174403}
}

@article{Yang2011-first-principles-PMA,
  title = {First-principles investigation of the very large perpendicular magnetic anisotropy at {Fe}$\textbar${MgO} and {Co}$\textbar${MgO} interfaces},
  author = {Yang, H. X. and Chshiev, M. and Dieny, B. and Lee, J. H. and Manchon, A. and Shin, K. H.},
  journal = {Phys. Rev. B},
  volume = {84},
  issue = {5},
  pages = {054401},
  numpages = {5},
  year = {2011},
  month = {Aug},
  publisher = {American Physical Society},
  doi = {10.1103/PhysRevB.84.054401},
  url = {https://link.aps.org/doi/10.1103/PhysRevB.84.054401}
}

@article{Niranjan2010-electric-field-effect-Fe-MgO,
    author = {Niranjan, Manish K. and Duan, Chun-Gang and Jaswal, Sitaram S. and Tsymbal, Evgeny Y.},
    title = {Electric field effect on magnetization at the {Fe/MgO}(001) interface},
    journal = {Appl. Phys. Lett.},
    volume = {96},
    xnumber = {22},
    pages = {222504},
    year = {2010},
    month = {06},
    issn = {0003-6951},
    doi = {10.1063/1.3443658},
    url = {https://doi.org/10.1063/1.3443658},
}

@Article{He2022-high-x-ray-dose-MgO-degradation,
author={He, Qi
and Shi, Hui
and Wang, Yinquan
and Cao, Lichao
and Gu, Xiang
and Wu, Jianwei
and Hong, Genshen
and Li, Minghua},
title={High-dose {X}-ray radiation induced {MgO} degradation and breakdown in spin transfer torque magnetic tunnel junctions},
journal={Sci. Rep.},
year={2022},
month={Nov},
day={03},
volume={12},
xnumber={1},
pages={18620},
issn={2045-2322},
doi={10.1038/s41598-022-19342-x},
url={https://doi.org/10.1038/s41598-022-19342-x}
}

@Article{Cheng2024-soft-X-ray-damage,
author={Cheng, Shang-Hong
and Chang, Chien-Hung
and Velasco-Velez, Juan-Jesus
and Liu, Bo-Hong},
title={Soft {X}-ray Induced Radiation Damage in Dip-and-Pull Photon Absorption and Photoelectron Emission Experiments},
journal={J. Phys. Chem. C},
year={2024},
month={Aug},
day={29},
publisher={American Chemical Society},
volume={128},
xnumber={34},
pages={14381-14387},
issn={1932-7447},
doi={10.1021/acs.jpcc.4c01067},
url={https://doi.org/10.1021/acs.jpcc.4c01067}
}

@article{Devolder2013-irradiation-induced-PMA-CoFeB-MgO,
    author = {Devolder, T. and Barisic, I. and Eimer, S. and Garcia, K. and Adam, J.-P. and Ockert, B. and Ravelosona, D.},
    title = {Irradiation-induced tailoring of the magnetism of {CoFeB/MgO} ultrathin films},
    journal = {J. Appl. Phys.},
    volume = {113},
    xnumber = {20},
    pages = {203912},
    year = {2013},
    month = {05},
    issn = {0021-8979},
    doi = {10.1063/1.4808102},
    url = {https://doi.org/10.1063/1.4808102},
}

@article{CHEN2010-effect-grain-size-CoFeB-films,
title = {Effect of grain size on magnetic and nanomechanical properties of {Co$_{60}$Fe$_{20}$B$_{20}$} thin films},
journal = {J. Alloys Compd.},
volume = {498},
xnumber = {2},
pages = {113-117},
year = {2010},
issn = {0925-8388},
doi = {https://doi.org/10.1016/j.jallcom.2010.03.141},
url = {https://www.sciencedirect.com/science/article/pii/S0925838810006468},
author = {Yuan-Tsung Chen and C.C. Chang}
}

@article{Chatterjee2019-physicochemical-origins-improvement-STT-RAM,
    author = {Chatterjee, Jyotirmoy and Gautier, Eric and Veillerot, Marc and Sousa, Ricardo C. and Auffret, Stéphane and Dieny, Bernard},
    title = {Physicochemical origin of improvement of magnetic and transport properties of {STT-MRAM} cells using tungsten on {F}e{C}o{B} storage layer},
    journal = {Appl. Phys. Lett.},
    volume = {114},
    xnumber = {9},
    pages = {092407},
    year = {2019},
    month = {03},
    issn = {0003-6951},
    doi = {10.1063/1.5081912},
    url = {https://doi.org/10.1063/1.5081912},
}

@ARTICLE{Capriata2022-grain-boundaries-SHNOs,
  author={Capriata, Corrado Carlo Maria and Jiang, Sheng and \AA{}kerman, Johan and Malm, Bengt Gunnar},
  journal={IEEE Electron Device Lett.}, 
  title={Impact of Random Grain Structure on Spin-{H}all Nano-Oscillator Modal Stability}, 
  year={2022},
  volume={43},
  xnumber={2},
  pages={312-315},
  doi={10.1109/LED.2021.3137952}}

@article{fulara2020natcomm,
	title = {Giant voltage-controlled modulation of spin {Hall} nano-oscillator damping},
	volume = {11},
	issn = {2041-1723},
	language = {en},
	xnumber = {1},
	urldate = {2020-08-12},
	journal = ntc,
	author = {Fulara, Himanshu and Zahedinejad, Mohammad and Khymyn, Roman and Dvornik, Mykola and Fukami, Shunsuke and Kanai, Shun and Ohno, Hideo and {\AA}kerman, Johan},
	month = dec,
	year = {2020},
	pages = {4006},
	URL = {https://doi.org/10.1038/s41467-020-17833-x}
}

@article{alexander2010magnonic,
  title={Magnonic logic circuits},
  author={Alexander, Khitun and Mingqiang, Bao and Kang, L Wang},
  journal={J. Phys. D: Appl. Phys},
  volume={43},
  xnumber={26},
  pages={264005},
  year={2010},
  doi = {http://stacks.iop.org/JPhysD/43/264005}
}

@article{khitun2011non,
  title={Non-volatile magnonic logic circuits engineering},
  author={Khitun, Alexander and Wang, Kang L},
  journal={J. Appl. Phys.},
  volume={110},
  xnumber={3},
  pages={034306},
  year={2011},
  publisher={American Institute of Physics},
  doi = {10.1063/1.3609062}
}

@article{behera2022energy,
  title = {Energy-Efficient {W}$_{100\ensuremath{-}x}${Ta}$_{x}/${C}o-{F}e-{B}/{M}g{O} Spin {H}all Nano-Oscillators},
  author = {Behera, Nilamani and Fulara, Himanshu and Bainsla, Lakhan and Kumar, Akash and Zahedinejad, Mohammad and Houshang, Afshin and \AA{}kerman, Johan},
  journal = {Phys. Rev. Appl.},
  volume = {18},
  issue = {2},
  pages = {024017},
  numpages = {8},
  year = {2022},
  month = {Aug},
  publisher = {American Physical Society},
  doi = {10.1103/PhysRevApplied.18.024017},
  url = {https://link.aps.org/doi/10.1103/PhysRevApplied.18.024017},
}

@article{zahedinejad2022memristive,
  title={Memristive control of mutual spin {H}all nano-oscillator synchronization for neuromorphic computing},
  author={Zahedinejad, Mohammad and Fulara, Himanshu and Khymyn, Roman and Houshang, Afshin and Dvornik, Mykola and Fukami, Shunsuke and Kanai, Shun and Ohno, Hideo and {\AA}kerman, Johan},
  journal={Nat. Mater.},
  volume={21},
  xnumber={1},
  pages={81--87},
  year={2022},
  publisher={Nature Publishing Group},
  doi = {s41563-021-01153-6}
}

@article{behera2024,
 title = {Ultra-Low Current 10 nm Spin Hall Nano-Oscillators},
 author = {Behera, Nilamani and Chaurasiya, Avinash Kumar and González, Victor H. and Litvinenko, Artem and Bainsla, Lakhan and Kumar, Akash and Khymyn, Roman and Awad, Ahmad A. and Fulara, Himanshu and \AA{}kerman, Johan},
 journal = {Advanced Materials},
 volume = {36},
 xnumber = {5},
 pages = {2305002},
 year = {2024},
 publisher={Wiley}
}

@article{kumar2023robust,
  title={Robust mutual synchronization in long spin {H}all nano-oscillator chains},
  author={Kumar, Akash and Fulara, Himanshu and Khymyn, Roman and Litvinenko, Artem and Zahedinejad, Mohammad and Rajabali, Mona and Zhao, Xiaotian and Behera, Nilamani and Houshang, Afshin and Awad, Ahmad A and others},
  journal={Nano Lett.},
  volume={23},
  xnumber={14},
  pages={6720--6726},
  year={2023},
  publisher={ACS Publications},
doi = {https://doi.org/10.1021/acs.nanolett.3c02036}
}

@article{rajabali2023injection,
  title={Injection locking of linearlike and soliton spin-wave modes in nanoconstriction spin {H}all nano-oscillators},
  author={Rajabali, Mona and Ovcharov, Roman and Khymyn, Roman and Fulara, Himanshu and Kumar, Akash and Litvinenko, Artem and Zahedinejad, Mohammad and Houshang, Afshin and Awad, Ahmad A and {\AA}kerman, Johan},
  journal={ Phys. Rev. Appl.},
  volume={19},
  xnumber={3},
  pages={034070},
  year={2023},
  publisher={APS}
}

@article{Gonzalez2024-spintronic-devices-next-generation-computation,
title = {Spintronic devices as next-generation computation accelerators},
journal = {Curr. Opin. Solid State Mater. Sci.},
volume = {31},
pages = {101173},
year = {2024},
issn = {1359-0286},
doi = {https://doi.org/10.1016/j.cossms.2024.101173},
url = {https://www.sciencedirect.com/science/article/pii/S1359028624000391},
author = {Victor H. González and Artem Litvinenko and Akash Kumar and Roman Khymyn and Johan \AA{}kerman}
}

@Article{Juge2021-He-ions-skyrmions,
author={Juge, Rom{\'e}o
and Bairagi, Kaushik
and Rana, Kumari Gaurav
and Vogel, Jan
and Sall, Mamour
and Mailly, Dominique
and Pham, Van Tuong
and Zhang, Qiang
and Sisodia, Naveen
and Foerster, Michael
and Aballe, Lucia
and Belmeguenai, Mohamed
and Roussign{\'e}, Yves
and Auffret, St{\'e}phane
and Buda-Prejbeanu, Liliana D.
and Gaudin, Gilles
and Ravelosona, Dafin{\'e}
and Boulle, Olivier},
title={Helium Ions Put Magnetic Skyrmions on the Track},
journal={Nano Lett.},
year={2021},
month={Apr},
day={14},
publisher={American Chemical Society},
volume={21},
xnumber={7},
pages={2989-2996},
issn={1530-6984},
doi={10.1021/acs.nanolett.1c00136},
url={https://doi.org/10.1021/acs.nanolett.1c00136}
}

@Article{Kern2022-he-ion-generation-and-guiding-skyrmions,
author={Kern, Lisa-Marie
and Pfau, Bastian
and Deinhart, Victor
and Schneider, Michael
and Klose, Christopher
and Gerlinger, Kathinka
and Wittrock, Steffen
and Engel, Dieter
and Will, Ingo
and G{\"u}nther, Christian M.
and Liefferink, Rein
and Mentink, Johan H.
and Wintz, Sebastian
and Weigand, Markus
and Huang, Meng-Jie
and Battistelli, Riccardo
and Metternich, Daniel
and B{\"u}ttner, Felix
and H{\"o}flich, Katja
and Eisebitt, Stefan},
title={Deterministic Generation and Guided Motion of Magnetic Skyrmions by Focused He+-Ion Irradiation},
journal={Nano Lett.},
year={2022},
month={May},
day={25},
publisher={American Chemical Society},
volume={22},
xnumber={10},
pages={4028-4035},
issn={1530-6984},
doi={10.1021/acs.nanolett.2c00670},
url={https://doi.org/10.1021/acs.nanolett.2c00670}
}

@article{Magnonics_Spin_Wave_Logic_Gates,
author = {Schneider, T. and Serga, A. A. and Leven, B. and Hillebrands, B. and Stamps, R. L. and Kostylev, M. P.},
doi = {10.1063/1.2834714},
issn = {00036951},
journal = {Appl. Phys. Lett.},
xnumber = {2},
pages = {022505},
title = {{Realization of spin-wave logic gates}},
url = {https://doi.org/10.1063/1.2834714},
volume = {92},
year = {2008}
}

@article{Gonzalez2024-global-biasing-SWIM,
    author = {González, Victor H. and Litvinenko, Artem and Khymyn, Roman and \AA{}kerman, Johan},
    title = {Global biasing using a hardware-based artificial {Z}eeman term in spinwave {I}sing machines},
    journal = {Appl. Phys. Lett.},
    volume = {124},
    xnumber = {9},
    pages = {092409},
    year = {2024},
    month = {02},
    issn = {0003-6951},
    doi = {10.1063/5.0185888},
    url = {https://doi.org/10.1063/5.0185888},
}

@Article{Litvinenko2023-SWIM,
author={Litvinenko, Artem
and Khymyn, Roman
and Gonz{\'a}lez, Victor H.
and Ovcharov, Roman
and Awad, Ahmad A.
and Tyberkevych, Vasyl
and Slavin, Andrei
and {\AA}kerman, Johan},
title={A spinwave {I}sing machine},
journal={Commun. Phys.},
year={2023},
month={Aug},
day={25},
volume={6},
xnumber={1},
pages={227},
issn={2399-3650},
doi={10.1038/s42005-023-01348-0},
url={https://doi.org/10.1038/s42005-023-01348-0}
}

@Article{Kumar2025-spin-wave-mediated-shno-sync,
author={Kumar, Akash
and Chaurasiya, Avinash Kumar
and Gonz{\'a}lez, Victor H.
and Behera, Nilamani
and Alem{\'a}n, Ademir
and Khymyn, Roman
and Awad, Ahmad A.
and {\AA}kerman, Johan},
title={Spin-wave-mediated mutual synchronization and phase tuning in spin Hall nano-oscillators},
journal={Nat. Phys.},
year={2025},
month={Feb},
day={01},
volume={21},
xnumber={2},
pages={245-252},
issn={1745-2481},
doi={10.1038/s41567-024-02728-1},
url={https://doi.org/10.1038/s41567-024-02728-1}
}

@article{mumax3,
    author = {Vansteenkiste, Arne and Leliaert, Jonathan and Dvornik, Mykola and Helsen, Mathias and Garcia-Sanchez, Felipe and Van Waeyenberge, Bartel},
    title = "{The design and verification of MuMax3}",
    journal = {AIP Advances},
    volume = {4},
    xnumber = {10},
    pages = {107133},
    year = {2014},
    month = {10},
    issn = {2158-3226},
    doi = {10.1063/1.4899186},
    url = {https://doi.org/10.1063/1.4899186},
}

@online{COMSOL,
   author    = "COMSOL",
    title     = "COMSOL multiphysics software",
    url       = "https://www.comsol.com/"
}

@article{Gonzalez2022,
    author = {González, Victor H. and Khymyn, Roman and Fulara, Himanshu and Awad, Ahmad A. and \AA{}kerman, Johan},
    title = "{Voltage control of frequency, effective damping, and threshold current in nano-constriction-based spin Hall nano-oscillators}",
    journal = {Appl. Phys. Lett.},
    volume = {121},
    xnumber = {25},
    pages = {252404},
    year = {2022},
    month = {12},
    issn = {0003-6951},
    doi = {10.1063/5.0128786},
    url = {https://doi.org/10.1063/5.0128786},
}

@article{Jiang2024review,
    author = {Jiang, Sheng and Yao, Linrong and Wang, Shun and Wang, Di and Liu, Long and Kumar, Akash and Awad, Ahmad A. and Litvinenko, Artem and Ahlberg, Martina and Khymyn, Roman and Chung, Sunjae and Xing, Guozhong and \AA{}kerman, Johan},
    title = {Spin-torque nano-oscillators and their applications},
    journal = {Appl. Phys. Rev.},
    volume = {11},
    xnumber = {4},
    pages = {041309},
    year = {2024},
    month = {10},
    issn = {1931-9401},
    doi = {10.1063/5.0221877},
    url = {https://doi.org/10.1063/5.0221877},
}

@article{Farle1998,
year = {1998},
publisher = {},
volume = {61},
xnumber = {7},
pages = {755},
author = {Michael Farle},
title = {Ferromagnetic resonance of ultrathin metallic layers},
journal = {Rep. Prog. Phys.}
}

@ARTICLE{Ahlberg2022,
author={Ahlberg, M. and Chung, S. and Jiang, S. and Frisk, A. and Khademi, M. and Khymyn, R. and Awad, A.~A. and Le, Q.~T. and Mazraati, H. and Mohseni, M. and Weigand, M. and Bykova, I. and Gro{\ss}, F. and Goering, E. and Sch\"{u}tz, G. and Gr\"{a}fe, J. and \AA{}kerman, J.},
title={Freezing and thawing magnetic droplet solitons},
journal={Nat. Commun.},
year={2022},
volume={13},
xnumber={1},
doi={10.1038/s41467-022-30055-7},
pages={2462}
}

@Article{Electromagnetic_Radiation_On_MTJs,
AUTHOR = {Seifu, Dereje and Peng, Qing and Sze, Kit and Hou, Jie and Gao, Fei and Lan, Yucheng},
TITLE = {Electromagnetic Radiation Effects on {MgO}-Based Magnetic Tunnel Junctions: A Review},
JOURNAL = {Molecules},
VOLUME = {28},
YEAR = {2023},
XNUMBER = {10},
PAGES = {4151},
URL = {https://www.mdpi.com/1420-3049/28/10/4151},
PubMedID = {37241892},
ISSN = {1420-3049},
DOI = {10.3390/molecules28104151}
}

@article{STT-MRAM2024,
  title={Spin-transfer torque magnetoresistive random access memory technology status and future directions},
  author={Worledge, Daniel C and Hu, Guohan},
  journal={Nat. Rev. Electr. Eng.},
  volume={1},
  xnumber={11},
  pages={730--747},
  year={2024},
  publisher={Nature Publishing Group UK London},
  url = {https://doi.org/10.1038/s44287-024-00111-z},
  doi = {10.1038/s44287-024-00111-z}
}

@article{gross2021miep,
  title={{MIEP}--{A} time-resolved {X}-ray image evaluation program},
  author={Gro{\ss}, Felix and Tr{\"a}ger, Nick and Gr{\"a}fe, Joachim},
  journal={SoftwareX},
  volume={15},
  pages={100705},
  year={2021},
  publisher={Elsevier}
}

@Article{Wittrock2024,
  author    = {Wittrock, Steffen and Perna, Salvatore and Lebrun, Romain and Ho, Katia and Dutra, Roberta and Ferreira, Ricardo and Bortolotti, Paolo and Serpico, Claudio and Cros, Vincent},
  journal   = {Nature Communications},
  title     = {Non-hermiticity in spintronics: oscillation death in coupled spintronic nano-oscillators through emerging exceptional points},
  year      = {2024},
  issn      = {2041-1723},
  month     = feb,
  number    = {1},
  volume    = {15},
  doi       = {10.1038/s41467-023-44436-z},
  publisher = {Springer Science and Business Media LLC},
}

@Article{Leutenantsmeyer2015,
  author    = {Johannes Christian Leutenantsmeyer and Vladyslav Zbarsky and Marvin von der Ehe and Steffen Wittrock and Patrick Peretzki and Henning Schuhmann and Andy Thomas and Karsten Rott and G{\"u}nter Reiss and Tae Hee Kim and Michael Seibt and Markus M{\"u}nzenberg},
  title     = {Spin-Transfer Torque Switching at Ultra Low Current Densities},
  journal   = {{MATERIALS} {TRANSACTIONS}},
  year      = {2015},
  volume    = {56},
  number    = {9},
  pages     = {1323--1326},
  doi       = {10.2320/matertrans.ma201570},
  publisher = {Japan Institute of Metals},
}

@article{Cheng2019prb,
  title = {Magnetic domain wall skyrmions},
  author = {Cheng, Ran and Li, Maxwell and Sapkota, Arjun and Rai, Anish and Pokhrel, Ashok and Mewes, Tim and Mewes, Claudia and Xiao, Di and De Graef, Marc and Sokalski, Vincent},
  journal = {Phys. Rev. B},
  volume = {99},
  issue = {18},
  pages = {184412},
  year = {2019},
  month = {May},
  publisher = {American Physical Society},
  doi = {10.1103/PhysRevB.99.184412},
  url = {https://link.aps.org/doi/10.1103/PhysRevB.99.184412}
}

@article{Jaiswal2017apl,
    author = {Jaiswal, S. and Litzius, K. and Lemesh, I. and B\"{u}ttner, F. and Finizio, S. and Raabe, J. and Weigand, M. and Lee, K. and Langer, J. and Ocker, B. and Jakob, G. and Beach, G. S. D. and Kl\"{a}ui, M.},
    title = {Investigation of the Dzyaloshinskii-Moriya interaction and room temperature skyrmions in W/CoFeB/MgO thin films and microwires},
    journal = {Appl. Phys. Lett.},
    volume = {111},
    number = {2},
    pages = {022409},
    year = {2017},
    month = {07},
    issn = {0003-6951},
    doi = {10.1063/1.4991360},
    url = {https://doi.org/10.1063/1.4991360}
}

@article{Tacchi2017prl,
  title = {Interfacial Dzyaloshinskii-Moriya Interaction in $\mathrm{Pt}/\mathrm{CoFeB}$ Films: Effect of the Heavy-Metal Thickness},
  author = {Tacchi, S. and Troncoso, R. E. and Ahlberg, M. and Gubbiotti, G. and Madami, M. and \AA{}kerman, J. and Landeros, P.},
  journal = {Phys. Rev. Lett.},
  volume = {118},
  issue = {14},
  pages = {147201},
  numpages = {6},
  year = {2017},
  month = {Apr},
  publisher = {American Physical Society},
  doi = {10.1103/PhysRevLett.118.147201},
  url = {https://link.aps.org/doi/10.1103/PhysRevLett.118.147201}
}

\end{document}